\documentstyle[11pt,aaspp4,flushrt]{article} 
\begin{document}

\title{The Spectra of Main Sequence Stars in Galactic Globular Clusters
I. CH and CN Bands in M13\altaffilmark{1}}

\author{Judith G. Cohen\altaffilmark{2}}

\altaffiltext{1}{Based on observations obtained at the
W.M. Keck Observatory, which is operated jointly by the California 
Institute of Technology and the University of California}
\altaffiltext{2}{Palomar Observatory, Mail Stop 105-24,
California Institute of Technology}

\begin{abstract}

Spectra with a high signal-to-noise ratio of 50 stars 
which are just below the main sequence
turnoff and are members of M13 are presented.  They yield
indices for the strength of the CH and the ultraviolet CN band.
There is no evidence for a variation in the strength of
either feature from star to star in this 
intermediate-metallicity galactic globular cluster,
and thus no evidence for primordial variations in the abundance
of C and N in M13.  This supports the hypothesis that abundance
variations found among the light elements in the evolved stars of M13
by Suntzeff (1981), and commonly seen on the giant and subgiant branches
of globular clusters of comparable metallicity, are due primarily
or entirely to mixing within a fraction of individual stars as they
evolve.

\end{abstract}

\subjectheadings{globular clusters: general --- globular clusters: individual (M13) --- stars: evolution}

\section{INTRODUCTION}

The galactic globular clusters contain some of the oldest stars
in our galaxy.  The main sequence in such a cluster
consists of the unevolved member stars still in their H-burning phases.
There is a long tradition of
research on the much brighter
globular cluster giants, for example high dispersion
spectroscopy by Cohen (1983), Pilachowski (1984) and Gratton \& Ortolani (1987)
among others, and of photometric studies (see, for example, Frogel, Cohen
\& Persson 1983).  These studies have contributed
much to our understanding of stellar evolution and
mixing processes, as well as of
the structure of
our Galaxy, galactic halos,
and the integrated light of old stellar systems.
A few pioneers have begun
to press downward in luminosity with high dispersion
spectroscopy reaching the horizontal
branch (Clementini {\it et al.} 1996,
Pilachowski {\it et al.} 1996, Cohen \& McCarthy 1997), 
but until recently no one has tried to reach
the globular cluster main sequence stars spectroscopically.

As reviewed by Kraft (1994) and by Briley, Hesser \& Smith (1994)
(see also the more general reviews by McWilliam 1997 and by Pinnsoneault 1997),
observational studies over the past two decades  
have revealed that variations of a factor of more than 10 for the
C, N, and O elements between globular cluster giants and sub-giants within
a particular globular cluster, including M13, are common.  Suntzeff (1981)
conducted the first detailed survey of the strength of the 
CH and CN bands among M13 giants.
The frequency of strong CN and weak CN stars among the red giants
is approximately equal, and the contrast in band strength 
between the two groups is large.
Weaker star-to-star variations of Na, Al, and Mg are often seen as well
among the giants in the best studied clusters.
These abundance variations obey specific
correlations; for example, enhanced N is accompanied by depleted carbon. 
Pilachowski {\it{et al.}} (1996) 
examined 130 giants and sub-giants in
M13 the faintest of which had M$_V$ = +1.2 mag,  slightly fainter
than the level of the horizontal branch.
They found star-to-star variations in Na abundance of a factor of 
6 and 
in Mg abundance of a factor of 4.  Sneden {\it{et al.}} (1997) explore the
behavior of C, N and O among the bright giants in M15, which is found to be similar to that shown by
giants in M13 and M92.  They find 
a star-to-star variation in the abundance of Eu and Ba in M15 which is
a factor of 4.  Similar variations among the giants
are detected within all globular clusters 
studied in suitable detail.  

Spectroscopic study of main sequence stars in galactic globular clusters 
is much more limited as the stars are faint, and achieving 
a suitable dispersion and signal-to-noise ratio
is difficult.  The only globular cluster whose
main sequence stars have been studied in some detail is 47 Tuc.
Hints of CN variations were found by Hesser (1978), and this work was
continued  by Hesser \& Bell (1980), Bell, Hesser \& Cannon (1983),
Briley, Hesser \& Bell (1991), 
Briley {\it{et al.}} (1994) and Briley {\it{et al.}} (1996).
The most recent work on 47 Tuc (Cannon {\it{et al.}} 1998)
demonstrates convincingly that variations in CN and CH are found at the
level of the main sequence.
Suntzeff (1989) and Suntzeff \& Smith (1991)
found very preliminary indications of an anti-correlation 
between CH and CN for
a small sample of main sequence stars in NGC 6752.
Molaro \& Pasquini (1994) got a relatively noisy spectrum of a single 
main sequence turnoff star in NGC 6397
to look for lithium.  Pilachowski \& Armandroff (1996) summed
spectra of 40 stars at the base of the giant branch in M13 in an unsuccessful
search for the (weak) 7700\AA\ OI triplet.  King, Stephens \& Boesgaard (1998)
attempted to derive [Fe/H] from Keck spectra of several subgiants near
the turnoff of M92, and got rather surprising results, [Fe/H] = $-$2.52,
a factor of two lower than the value in common use.

It was not believed possible to produce Na and Al in 
globular cluster red giants, given their low masses and
relatively unevolved evolutionary state, until
Denisenkov \& Denisenkova's (1990) seminal paper.  They
suggest that instead of neutron captures on Ne$^{22}$, 
proton captures on Ne$^{22}$ could produce Na and Al enhancements.
Since there is no source of free neutrons in globular cluster
red giants, the
former cannot occur in H-burning stars, while
the latter can.
(See also Langer, Hoffman \& Sneden 1993.) 
Debate still centers on
mixing versus primordial variations as the origin of
the observed abundance variations.  The interaction
between internal rotation and mixing may be a critical one (Sweigart 
\& Mengel 1979, Sweigart 1997), as may that between rotation and
diffusion in the photosphere.  Mass transfer in binary stars may
also play a role under some circumstances (McClure 1984, 1997).
Recently as a result of this understanding of how
Na and Al could be produced in the interiors of red giants and
mixed to the stellar surface via convection zones, the pendulum has swung
towards favoring mixing as the explanation for most of the observed
variations.  The current theoretical picture is summarized in Cavallo,
Sweigart \& Bell (1996), Langer, Hoffman \& Zaidens (1997)
and in Cannon {\it{et al.}} (1998).

Globular cluster main sequence stars should
not yet have synthesized through internal nuclear burning any elements
heavier than He (and Li and Be) and hence will
be essentially unpolluted by the internal nuclear burning 
and production of various heavy elements
that occur in later
stages of stellar evolution.  Theory predicts that these stars are
unaffected by gravitational settling and that their surfaces should
be a fair representation of the gas from which the globular cluster formed.
Thus the persistence of variations
in C and N 
to such low luminosities in 47 Tuc (Cannon {\it{et al.}} 1998 and
references therein) is surprising.

The advent of the Keck Telescope with the Low Resolution Imaging Spectrograph
(henceforth LRIS, Oke {\it{et al.}} 1995), an efficient multi-object spectrograph
coupled to a 10--m telescope, makes a
high precision
study of the spectra of main sequence stars in galactic globular
clusters feasible.
The major issue I intend to explore in this series of papers
is that of star-to-star
variations in abundances within a single globular cluster
at and below the level of the main sequence turnoff.

\section{THE SAMPLE OF STARS}

M13 was chosen to
begin this effort because it is nearby, hence the
turnoff stars will be relatively bright, it has very low
reddening, and its giants and sub-giants have been the
subject of an extensive series of papers by Kraft and his collaborators.
(See Pilachowski {\it{et al.}} 1996 for references to their earlier papers.)
Its high galactic latitude guarantees minimum contamination of a sample
by field stars.

Short exposure images in $B$ and $R$ were taken with LRIS centered 
on the field used
by Fahlman \& Richer (1986) in their photometric study of the
main sequence of M13.
Photometry was obtained 
with DAOPHOT (Stetson 1987) using these short exposures
calibrated on the system of Landolt (1992).
The zero point for each color in each field is uncertain by $\pm0.05$ mag.
A sample of main sequence stars was chosen based on their position
on the locus of the main sequence as defined by this photometry.
Each candidate was inspected for crowding and stars were chosen
for the spectroscopic sample on the
basis of minimum crowding.  A second field was chosen 400'' North
of this field, somewhat closer to the center of M13, and a similar
procedure was carried out. Table 1 gives the object's coordinates (J2000), 
$R$ mag,
$B-R$ color, and indices (together with their errors) for two molecular bands
for the M13 main sequence stars in the spectroscopic sample.

Since the fields are very crowded, in addition to providing
the star coordinates, we provide an identification chart for
a few stars in each of the two fields, from which, given the
accurate relative coordinates, the rest of the stars can be located.
Relative stellar coordinates are defined from the LRIS images themselves
assuming the telescope pointing recorded in the image header is correct
and taking into account the optical distortions in the images.  The
astrometry of Cudworth \& Monet (1979) is used to fix the
absolute coordinates.
%
Star M13ms J1641019+362403 in field 1 is the star at location
(141,273) in Figure 1 of Fahlman \& Richer (1986).
A finding chart for several stars in our sample in Field 2 is 
given in Figure 1 below.  

%
%
%

Figure 2 presents a
color-magnitude diagram for the main sequence stars in the M13 sample.  
The stars that have been observed spectroscopically are shown
as filled circles.  To guide the eye in establishing the
iosochrone locus, stars in Field 2 that lie in the region somewhat 
brighter and somewhat fainter than that spanned by the sample stars 
are shown as open circles; for the fainter stars, every eighth star
is plotted.

%

\section{SPECTROSCOPIC OBSERVATIONS AND MEASUREMENT OF BAND INDICES}

Two slitmasks, one in each field, were 
designed containing 50 stars from the M13 main sequence
star sample.  These were used at relatively low dispersion
with the LRIS (300 g/mm grating, 2.46\AA/pixel, 0.7 arcsec slit width) for
a spectral resolution of 8\AA.  The CCD detector is digitized
at 2 electrons/DN with a readout noise of 8 electrons.
Two 800 sec exposures were
taken with each slitmask under conditions of good seeing and
dark sky in the spring of 1998.  The data were reduced in
a straightforward manner as described in Cohen {\it{et al.}} (1999)
using Figaro (Shortridge 1988) except that the wavelength calibration came
from arc lamp exposures, rather than from night sky lines on the
spectra themselves.
The spectra are not fluxed.

All 50 stars are members of M13 based on the metal poor appearance
of their spectra and on their radial velocities.

Since these stars are metal poor, the absorption lines are in general quite
weak, and rather than adhere to the usual definition of a single
sided band index, we measure a CH index using
continuum bandpasses on both sides of the G band at 4300\AA, with
a feature bandpass chosen to avoid H$\gamma$.  Thus the
blue continuum bandpass goes from 4180 to 4250\AA, and the red one from
4380 to 4460 A.  The feature bandpass covers the wavelengths
4285 to 4315\AA.  Weights of 0.6 and 0.4 are assigned to the blue
and red continuum bandpasses respectively based on their offset
from the wavelength of the G band.
The CH index thus measured is given
in Table 1.  The values are the fraction of 
absorption from the continuum,
and are not in magnitudes.

For the ultraviolet CN band with its head at 3883\AA, because of
crowding by the higher Balmer lines,
it is impossible to find a suitable blue continuum bandpass.  Thus
the feature is defined in the usual way by a red continuum bandpass at 3894 to 3910\AA,
with the feature bandpass including 3860 to 3888\AA.  Again the 
index feature strengths as a fraction of absorption from the continuum are given
in Table 1.  A minimum continuum strength (700 DN/pixel) was established for
an accurate measurement of the uvCN index; 44 of the 50 stars
in the sample met this requirement, and no uvCN index is listed
for the six stars whose spectra did not achieve the necessary continuum level.
These are among the faintest stars, but not the six faintest, as slitmask
alignment also plays a role here, particularly for such narrow slits.

Errors (1$\sigma$) for the molecular band indices were calculated
based on Poisson statistics 
from the observed count rates in the continuum and in the feature bandpasses.
These are listed for each star in Table 1.  
The values given in Table 1 thus do
not include the effect of cosmic rays nor the effect of the 
background signal from the night sky, both of which are small.

Even if a mean continuum
were applied to normalize the spectra, the
observed dispersions within the defined continuum bandpasses
cannot themselves be used due to the probable presence of many weak absorption
features in the spectra.  Errors calculated in this way (without
normalizing the continuum) are typically
twice those calculated from the Poisson statistics, and provide
a firm upper limit on the uncertainties of the measured molecular
band indices.

To illustrate the quality of the spectra, Figure 3 shows the
spectrum of the brightest star at $R$ in our sample of main
sequence stars in M13 (M13ms J1641079+362413) and of the faintest star for
which both a CH index and a uvCN index were measured (M13ms J1641085+362317).
The thin line is the spectrum of the fainter star multiplied by a factor of 3.1
in the region around the CH band to facilitate comparison with the spectrum
of the brightest star.

%
%

\section{ANALYSIS}

The color-magnitude diagram for our sample of main sequence
stars in M13 displayed in Figure 2 illustrates a very important point.  The
total range in $(B-R)$ color of the sample members is very small,
$<0.2$ mag. The locus of the main sequence is very tight in 
color at a fixed luminosity.  The total range in
$R$ magnitude is less than 1.5 mag.  We therefore ignore
all subtleties, all model atmosphere and line synthesis
calculations, all calculations of molecular equilibria,
all variations in $T_{eff}$ and surface gravity,
and instead proceed in a very simple manner.

We rank the stars according to the position along the main
sequence, a ranking which is almost identical to that in $R$ magnitude,
with the star highest up the isochrone as first, and the star
lowest on the main sequence as having ``star order'' = 50.

Figures 4 and 5 show the results for the 50 M13 main sequence
stars.  (Only 44 stars are plotted in Figure 5 due to the 
minimum continuum level requirement
imposed for measuring a uvCN index.)  The 1$\sigma$
errors for each star are plotted as well.  A linear fit was derived
for the uvCN index, and the 1$\sigma$ rms dispersion around that
fit is 0.023 (2.3\% of the continuum), while the Poisson errors calculated from the
measurements range from 0.008 to 0.017.  For the CH index,
a quadratic fit was made, as shown in Figure 5.  The 
1$\sigma$ rms dispersion around that
fit is 0.010, while the calculated uncertainties for the faintest stars
are $\sim0.005$. 

It seems reasonable to assume that the rise in the strength of the
CH band index seen in Figure 5 for the most luminous stars in this sample is
due to their slightly redder color, hence slightly cooler $T_{eff}$
as one begins to turn off the main sequence (see Figure 2) enhancing
the strength of the CH band.  Because the CN index is defined as a single
sideband index with the continuum to the red of the feature, the
measured index is affected by the continuum slope as well as by
the feature strength.
The contribution from the
continuum itself to $I(uvCN)$ is on average $\sim$0.09.  There is
also some contribution to $I(uvCN)$ from the Balmer line H8 at
3889\AA, which in such metal poor stars may be a substantial fraction
of the total absorption within the bandpass of the uv CN feature.

\section{DISCUSSION}

Our data provide no evidence for variations of CH or CN band strengths
among the 50 main sequence stars in our sample in M13.  The
errors are small compared to the size of the measured indices. Our analysis
is extremely simple and does not depend in any way on model atmospheres
or spectral synthesis.  The CH bands are quite weak, and
hence one might argue that they hide a variation in band strength 
from star-to-star
which goes from $\epsilon$ to a maximum of $\sim6\%$, while still
having a range of more than a factor of two.
But a careful examination
of figure 5 makes this difficult to envision, as only one star has
a feature strength below 1\% (M13ms J1641072+362756, 
with $I(CH) = 0.008 \pm0.004$), 
and the errors are quite small for the entire sample.  The mean absorption
in the uvCN band is much larger, so such an argument cannot be
applied to $I(uvCN)$.  The
uvCN band indices in Figure 4 demonstrate
quite convincingly that in M13 there
are no star-to-star variations in the strength of this molecular band.
But here the issue of the contribution of H8 to this index in such
metal poor stars cannot be ignored.  A full spectral synthesis is
required to establish the maximum size of the variations that
can be hidden within the constraints of our data.  This is deferred
to a future paper in this series.

It is not clear how our results can be reconciled with the observations
in 47 Tuc of Cannon {\it{et al.}} (1998) and references therein,
where substantial
band strength variations were seen for main sequence stars.  Our sample
in M13 is larger and our spectra are of higher signal-to-noise ratio,
but M13 is somewhat farther away, the main sequence stars are somewhat
fainter than those in 47 Tuc, and the abundance of the cluster is lower.
The issue of membership for our sample of 50 main sequence stars is clear;
all are cluster members.

Similar data from the LRIS at the Keck Observatory
for an even larger sample of main sequence stars in M71 are now in hand
and the analysis and results of that sample should prove illuminating in 
trying to understand the origin of this discrepancy.

\section{SUMMARY}

I have determined the strength of the CH and CN bands 
from spectra of 50 main sequence stars in M13. 
The data would seem to suggest that large variations of
C and N are not seen at the level of the main sequence and below
it, but the reader is cautioned that a firm conclusion must await
a detailed analysis of C and N abundances which will appear in a
later paper (Briley \& Cohen 1999).
Significant primoridal variations of C and N do not 
appear to be present in M13.
This supports the hypothesis that abundance
variations found among the light elements in the evolved stars of M13
by Sneden (1981), and commonly seen on the giant and subgiant branches
of globular clusters of comparable metallicity, are due primarily
or entirely to mixing within a fraction of individual stars as they
evolve.

\acknowledgements The entire Keck/LRIS user community owes a huge debt
to Jerry Nelson, Gerry Smith, Bev Oke, and many other people who have
worked to make the Keck Telescope and LRIS a reality and to
operate and maintain the Keck Observatory.  We are grateful
to the W. M. Keck Foundation, and particularly its late president,
Howard Keck, for the vision to fund the construction of the W. M. Keck
Observatory.   I also thank Jim Hesser for a guide to the
literature on 47 Tuc, Kevin Richberg for help with the data
reduction and Patrick Shopbell for help at the telescope.

\newpage

\begin{deluxetable}{lrrrrrr}
\tablewidth{0pt}
\scriptsize
\tablecaption{Properties of the Sample of Main Sequence Stars in M13}
\tablehead{
\colhead{ID}  & \colhead {$R$} & 
\colhead{$(B-R)$} & \colhead{$I(uvCN)$} & \colhead{$\sigma$(uvCN)}
& \colhead{$I(CH)$} & \colhead{$\sigma$(CH)} \nl
 &  \colhead{(mag)} &
\colhead{(mag)} & \colhead{(\%)} & \colhead{(\%)} & 
\colhead{(\%)} & \colhead{(\%)} \nl
}
\startdata
%
%
M13ms J1641010+362529  &  18.69 & 0.77 & ... & ... &       0.047 &   0.003     \nl
M13ms J1641030+362529  &  18.20 & 0.73 &  0.202 &   0.011 & 0.021 &   0.002     \nl
M13ms J1641064+362508  &  17.84 & 0.85 & 0.228 &   0.011 & 0.018 &   0.002     \nl
M13ms J1641050+362454  &  18.50 & 0.74 & 0.181 &   0.010 & 0.026 &   0.002     \nl
M13ms J1641942+362438  &  18.54 & 0.74 & 0.221 &   0.012 & 0.028 &   0.002     \nl
M13ms J1641071+362431  &  18.47 & 0.75 & 0.229 &   0.011 & 0.029 &   0.002     \nl
M13ms J1641019+362403  &  18.67 & 0.76 & ... & ... &       0.037 &   0.003     \nl 
M13ms J1641079+362413  &  17.40 & 0.90 & 0.198 &   0.008 & 0.041 &   0.003     \nl
M13ms J1641065+362340  &  18.47 & 0.75 & ... & ... &       0.027 &   0.002     \nl
M13ms J1641058+362332  &  17.82 & 0.85 & 0.216 &   0.010 & 0.028 &   0.002     \nl
M13ms J1641080+362324  &  18.30 & 0.74 & 0.226 &   0.012 & 0.032 &   0.002     \nl
M13ms J1641085+362317  &  18.67 & 0.74 & 0.243 &   0.013 & 0.022 &   0.002     \nl
M13ms J1641070+362251  &  17.95 & 0.74 & 0.227 &   0.012 & 0.030 &   0.001     \nl
M13ms J1641068+362224  &  17.48 & 0.88 & 0.203 &   0.009 & 0.070 &   0.003     \nl
M13ms J1641076+362151  &  18.06 & 0.73 & 0.241 &   0.012 & 0.033 &   0.002     \nl
M13ms J1641099+362158  &  17.88 & 0.75 & 0.235 &   0.013 & 0.032 &   0.002     \nl
M13ms J1641022+362121  &  17.85 & 0.75 & 0.215 &   0.010 & 0.040 &   0.003     \nl
M13ms J1641076+362138  &  18.76 & 0.78 & 0.205 &   0.014 & 0.049 &   0.002     \nl
M13ms J1641031+362028  &  17.81 & 0.86 & 0.226 &   0.009 & 0.030 &   0.002     \nl
M13ms J1641012+362017  &  17.60 & 0.82 & 0.182 &   0.009 & 0.051 &   0.002     \nl
M13ms J1641023+361953  &  18.77 & 0.79 &  ... & ... &       0.047 &   0.003     \nl
M13ms J1641001+361937  &  18.30 & 0.78 & 0.228 &   0.009 & 0.043 &   0.002     \nl
M13ms J1641055+361909  &  18.50 & 0.73 & 0.228 &   0.010 & 0.018 &   0.002     \nl
M13ms J1641120+361926  &  18.59 & 0.74 & 0.224 &   0.010 & 0.029 &   0.002     \nl
M13ms J1641044+361836  &  17.57 & 0.85 & 0.167 &   0.005 & 0.065 &   0.003     \nl
M13ms J1641129+363226  &  18.15 & 0.73 &   0.248 &   0.013 & 0.032 &   0.003     \nl
M13ms J1641043+363218  &  18.19 & 0.73 &   0.245 &   0.010 & 0.025 &   0.002     \nl
M13ms J1641107+363152  &  18.18 & 0.71 &   0.242 &   0.013 & 0.024 &   0.002     \nl
M13ms J1641087+363124  &  18.06 & 0.73 &   0.231 &   0.017 & 0.038 &   0.002     \nl
M13ms J1641101+363112  &  17.67 & 0.83 &   0.193 &   0.013 & 0.052 &   0.003     \nl
M13ms J1641105+363058  &  17.75 & 0.82 &   0.209 &   0.016 & 0.056 &   0.004     \nl
M13ms J1641059+363050  &  18.17 & 0.73 &   0.265 &   0.013 & 0.033 &   0.002     \nl
M13ms J1641055+363041  &  17.95 & 0.75 &   0.256 &   0.010 & 0.040 &   0.002     \nl
M13ms J1641099+363027  &  18.19 & 0.72 &   0.230 &   0.017 & 0.040 &   0.003     \nl
M13ms J1641077+363011  &  18.22 & 0.76 & ... & ... &  0.034 &   0.002     \nl
M13ms J1641059+363000  &  17.68 & 0.84 &   0.198 &   0.013 & 0.073 &   0.004     \nl
M13ms J1641084+362943  &  17.71 & 0.83 &   0.217 &   0.011 & 0.039 &   0.003     \nl
M13ms J1641110+362905  &  17.62 & 0.85 &  0.217 &   0.013 & 0.052 &   0.004     \nl
M13ms J1641018+362821  &  17.75 & 0.80 &   0.222 &   0.012 & 0.057 &   0.003     \nl
M13ms J1641072+362756  &  18.23 & 0.73 &   0.278 &   0.013 & 0.008 &   0.004     \nl
M13ms J1641078+362746  &  18.00 & 0.74 &   0.248 &   0.009 & 0.034 &   0.002     \nl
M13ms J1641063+362732  &  18.25 & 0.72 &   0.228 &   0.015 & 0.031 &   0.002     \nl
M13ms J1641056+362724  &  18.19 & 0.76 &   0.299 &   0.014 & 0.026 &   0.002     \nl
M13ms J1641028+362705  &  17.77 &  0.80 &  0.192 &   0.008 & 0.058 &   0.003     \nl
M13ms J1641079+362645  &  18.15 & 0.72 &   0.243 &   0.010 & 0.020 &   0.002     \nl
M13ms J1641077+362628  &  17.90 & 0.76 &   0.216 &   0.013 & 0.027 &   0.003     \nl
M13ms J1641068+362545  &  18.07 & 0.73 &   0.227 &   0.010 & 0.031 &   0.003     \nl
M13ms J1641148+362833  &  17.64 &  0.84 &  0.194 &   0.018 & 0.058 &   0.003     \nl
M13ms J1640581+362618  &  17.76 & 0.80 & ... & ...   & 0.066 &   0.003     \nl
M13ms J1641120+363137  &  18.09 & 0.72 &   0.243 &   0.012 & 0.012 &   0.002     \nl
\enddata
\end{deluxetable}

\newpage

\begin{figure}
\epsscale{0.7}
\plotone{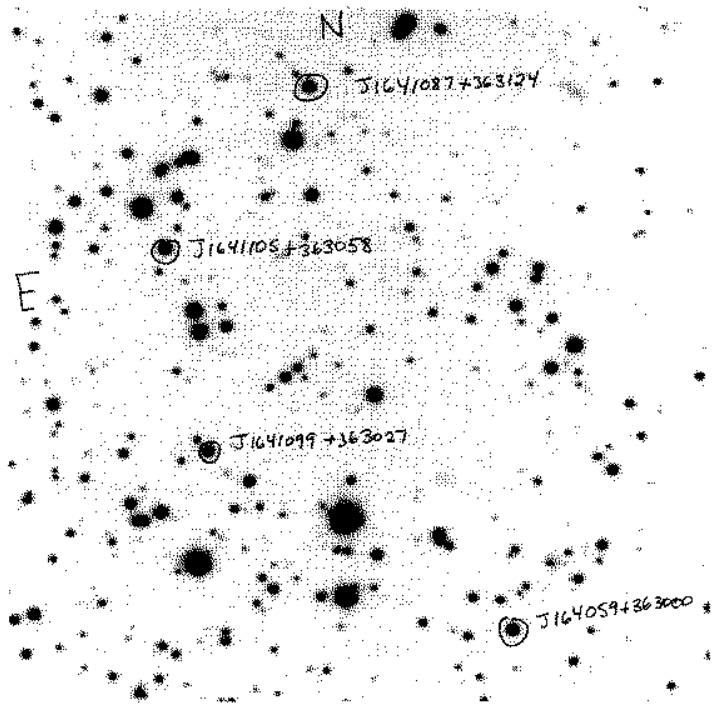}
\caption[jgcohen1.fig1.ps]{A square section 110 arcsec on a side from
a 15 sec $R$ exposure taken with LRIS of field 2 in M13 is shown. 
The locations of several M13 main sequence
stars in our sample in this field are indicated.\label{fig1}}
\end{figure}

\begin{figure}
\epsscale{0.7}
\plotone{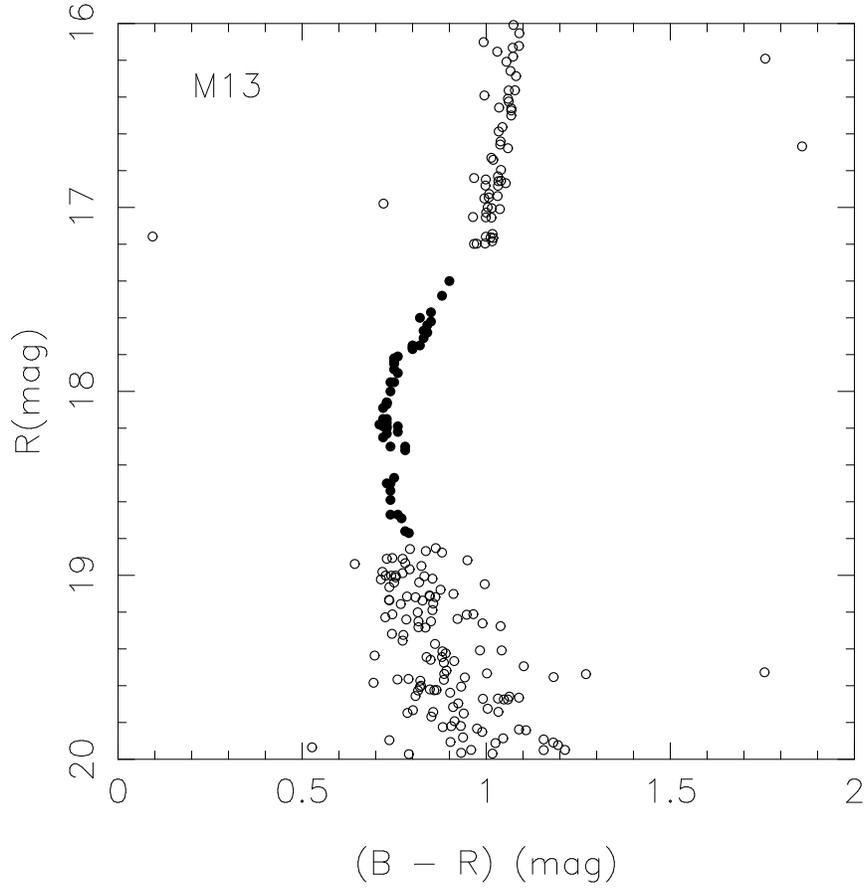}
\caption[jgcohen1.fig2.ps]{The color-magnitude diagram for the region of the
main sequence in M13.  The stars in our spectroscopic sample are shown
as filled circles.  Open circles are used to denote the stars in Field 2
in a region somewhat brighter than the spectroscopic sample. Every eighth star
in Field 2 is plotted as an open circle in a region somewhat
fainter than our main sequence sample. \label{fig2}}
\end{figure}

\begin{figure}
\epsscale{0.7}
\plotone{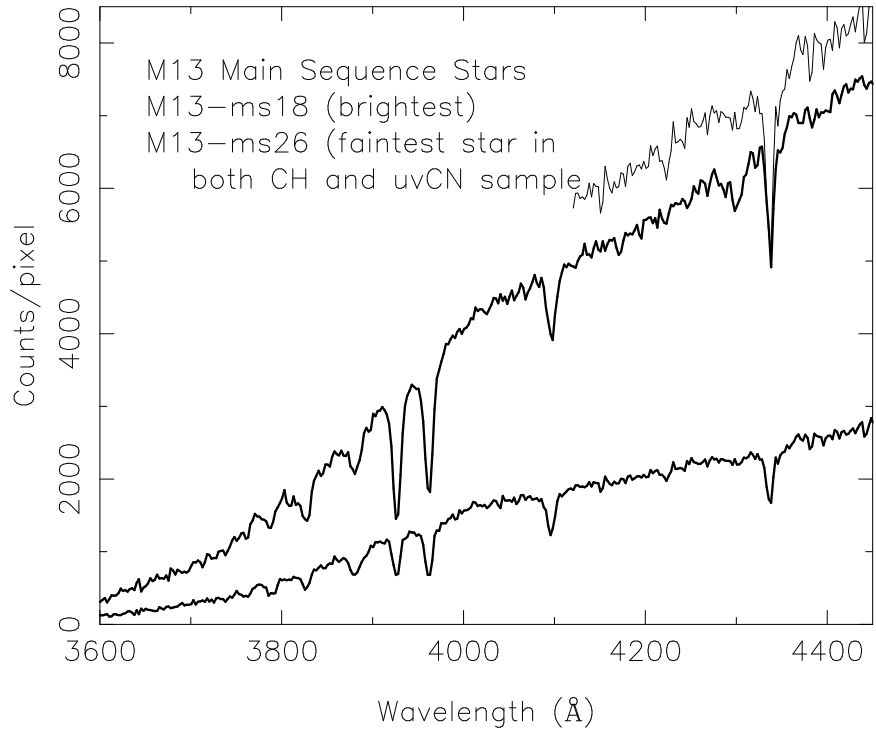}
\caption[jgcohen1.fig3.ps]{ Spectra of the brightest star in our sample
of main sequence stars in M13 (M13ms J1641079+362413) and of the faintest star
for which both CH and uvCN indices were measured (M13ms J1641085+362317) are shown
as thick solid lines.  The thin line shows the spectrum of the fainter star
in the region of the CH band multiplied by a factor of 3.1.
\label{fig3}}
\end{figure}

\begin{figure}
\epsscale{0.7}
\plotone{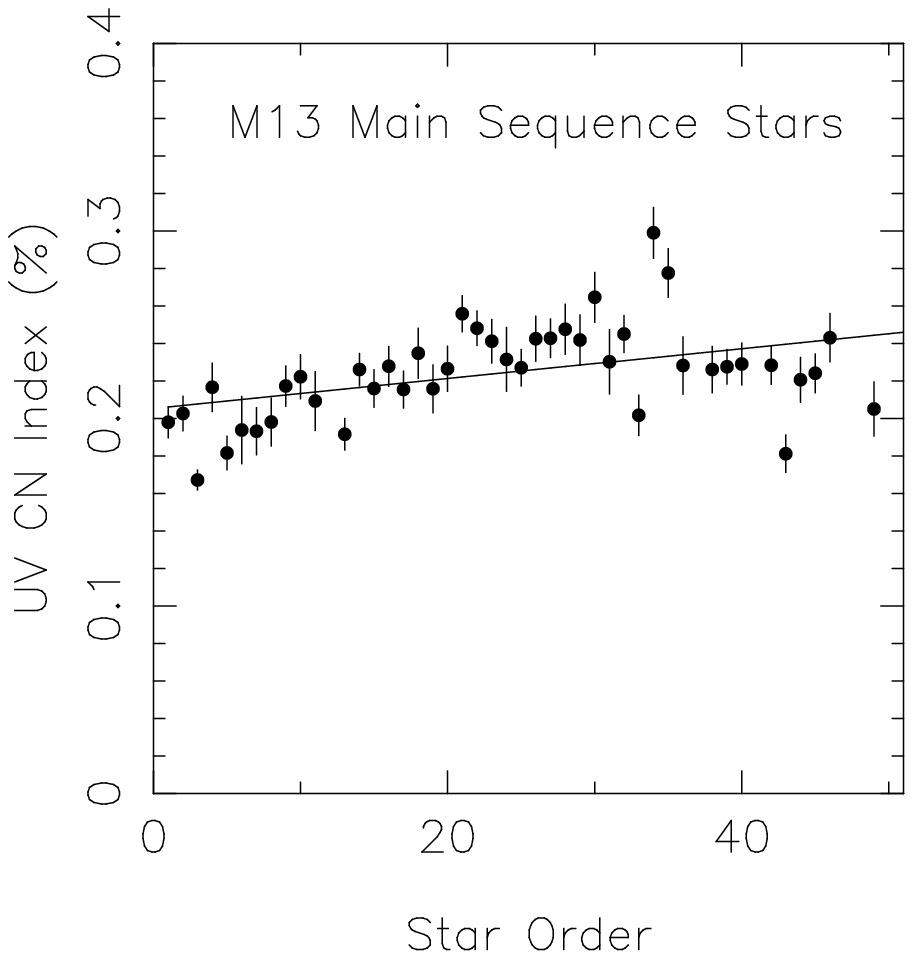}
\caption[jgcohen1.fig4.ps]{The uvCN indices for 44 main sequence stars in M13
are plotted as a function of star order (equivalent to position along
the main sequence, with the brightest star having ``Star Order'' = 1.
The error bars shown for each point are $1\sigma$ errors calculated from
the observed count rates assuming Poisson statistics.  The line is
a least squares fit to the observed points.
\label{fig4}}
\end{figure}

\begin{figure}
\epsscale{0.7}
\plotone{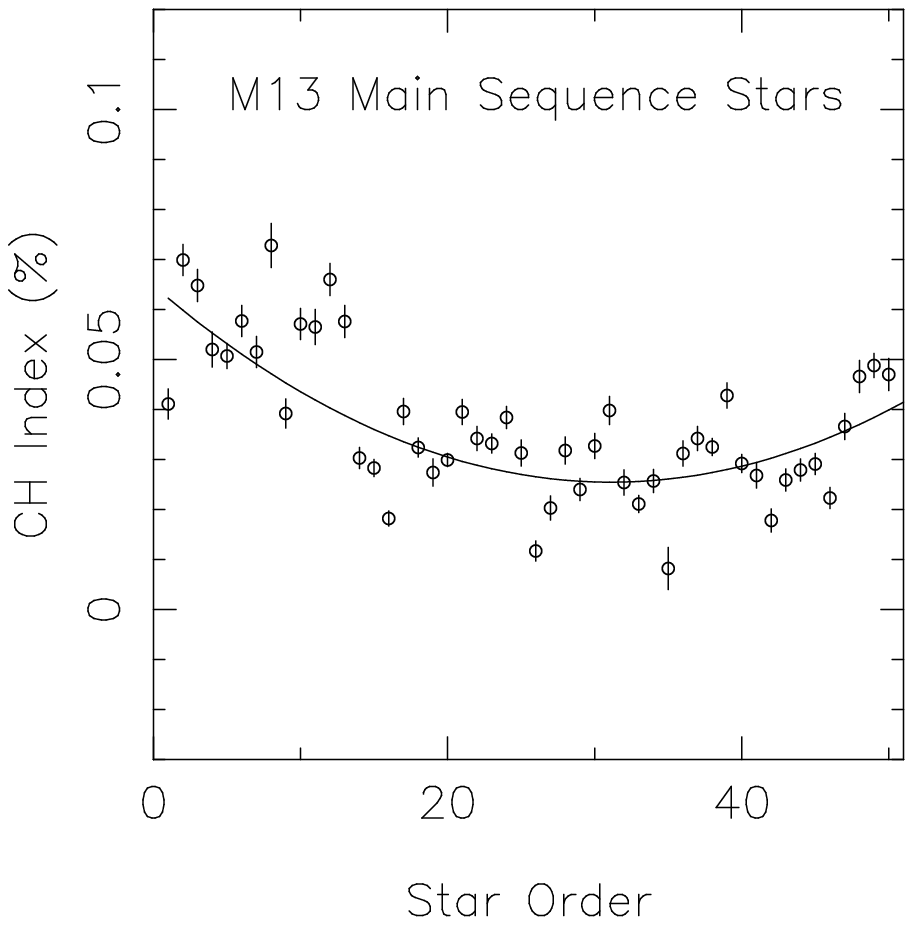}
\caption[jgcohen1.fig5.ps]{The G band CH indices for 50 main sequence stars in M13
are plotted as a function of star order (equivalent to position along
the main sequence, with the brightest star having ``Star Order'' = 1.
The error bars shown for each point are $1\sigma$ errors calculated from
the observed count rates assuming Poisson statistics.  The line is
a second order fit to the observed points.
\label{fig5}}
\end{figure}

\end{document}